\documentclass[12pt,preprint]{aastex}

\def\a4{\hsize 17.0cm \vsize 25.cm}

\shorttitle{superwinds \& further star formation}
\shortauthors{Silich et al.}

\begin{document}

\title{On the feedback from super stellar clusters. I. The structure of giant HII regions and HII galaxies}

\author{
Guillermo Tenorio-Tagle}
\affil{Instituto Nacional de Astrof\'\i sica Optica y
Electr\'onica, AP 51, 72000 Puebla, M\'exico; gtt@inaoep.mx}

\author{Casiana Mu\~noz-Tu\~n\'on}
\affil{Instituto de Astrof\'{\i}sica de Canarias, E 38200 La
Laguna, Tenerife, Spain; cmt@ll.iac.es}

\author{
Enrique P\'erez}
\affil{Instituto de Astrof\'{\i}sica de Andaluc\'\i{}a (CSIC), 
Camino bajo de Huetor 50, E 18080 Granada, Spain; eperez@iaa.es}

\author{
Sergiy Silich}
\affil{Instituto Nacional de Astrof\'\i sica Optica y
Electr\'onica, AP 51, 72000 Puebla, M\'exico; silich@inaoep.mx}
\affil{Visiting professor in the Department of Physics and
Astronomy at the University of Kentucky, Lexington, KY 40506-0055;
silich@pa.uky.edu}

\author{
Eduardo Telles}
\affil{Observat\'orio Nacional, 
Rua Jos\'e Cristino 77, 20921-400, Rio de Janeiro, Brazil; etelles@on.br}

\begin{abstract}

We review the structural properties of giant extragalactic HII regions and 
HII galaxies based on  two dimensional hydrodynamic calculations, and
propose an evolutionary sequence that accounts for their observed detailed 
structure. The model assumes a massive and young stellar cluster surrounded 
by a large collection of clouds. These are thus exposed to the  most 
important star-formation feedback mechanisms:  photoionization and the 
cluster wind. The models show how the two feedback mechanisms compete with 
each other in the disruption of clouds and lead to two different hydrodynamic 
solutions: The storage of clouds into a long lasting ragged shell that 
inhibits the expansion of the thermalized wind, and the steady filtering of 
the shocked wind gas through channels carved within the cloud stratum. Both 
solutions are here claimed to be concurrently at work in giant HII regions 
and HII galaxies, causing their detailed inner structure. This includes 
multiple large-scale shells, filled with an X-ray emitting gas, that evolve 
to finally merge with each other, giving the appearance of shells within 
shells. The models also show how  the  inner  filamentary structure of the 
giant superbubbles is largely enhanced with matter ablated from clouds and 
how cloud ablation proceeds within the original cloud stratum. The 
calculations point at the initial contrast density between the cloud and 
the intercloud media as the factor that defines which of the two feedback 
mechanisms becomes dominant throughout the evolution. Animated version of 
the models presented can be found at 
http://www.iaa.csic.es/\~{}eperez/ssc/ssc.html.


\end{abstract}

\section{Introduction}

Multiple studies during the last decades have addressed the impact of photoionization, stellar winds and supernova explosions 
on interstellar matter (ISM). Disruptive events that re-structure the birth place of massive stars and their surroundings,
while leading to multiple  phase transitions in the ISM (see e.g. Comeron 1997; Yorke et al. 1989; 
Garc\'i{}a-Segura et al. 2004; Hosokawa \& Inutsuka 2005 and  Tenorio-Tagle \& Bodenheimer 1988 and references therein). On the other hand,
giant molecular clouds, the sites of ongoing star formation, present a hierarchy of clumps and filaments of different scales
whose volume filling factor varies between  10$\%$ to 0.1$\%$ (see McLow \& Klessen, 2004, and references therein). 
As the efficiency of star formation in molecular clouds is estimated to be $\leq $ 10$\%$ (Larson 1988; Franco et al. 1994) the implication is that 
the bulk of the cloud structure remains after a star forming episode.
All of these studies are relevant within the fields of interstellar matter and star formation 
and, in particular, regarding  the physics of feedback, a major ingredient in the evolution of galaxies and thus in cosmology.

From the observations, we know that the energetics of major bursts of star formation allow us to track galaxies up to very
high redshifts, however,  by the time these sources are found star formation is already in a well advanced stage of its evolution,
and thus we do not know for certain, not even for the closest sources, which is the state of the matter left over from a major massive  burst of 
star formation. We do know that
giant extragalactic HII regions and HII galaxies are excellent examples of the impact of massive stars on the ISM. They  
belong to the same class of objects because  they are powered by 
young  massive bursts of star formation. Also because of their similar physical size, morphology and inner structure. 
Detailed studies of 30 Doradus (see Chu \& Kennicutt 1994; Melnick et al. 1999), NGC 604 (Sabalisck et al. 1995; Yang et al 1996) and 
several giant extragalactic HII regions and HII galaxies (Mu\~noz-Tu\~n\'on et al. 1996; 
Telles et al. 2001; Ma\'\i z-Apell\'aniz et al. 1999)
have been designed with the aim of unveiling the inner structure and dynamics of the nearest 
examples.
All of them present a collection of nested shells that enclose an X-ray emmiting gas and that 
may extend up to kpc scales. Some of the largest shells 
have stalled while others present 
expansion speeds of up to several tens of km s$^{-1}$. Detailed HST images
have also confirmed these issues  for HII galaxies as well (Martin et al. 2002; http://hubblesite.org/newscenter/newsdesk/archive/releases/2004/35/image/a). 
All of these sources, as in the studies of Telles et al (2001) and Cair\'os et al. (2001), present  
a central  bright condensation coincident with the massive burst of stellar formation. 
In giant HII regions and in some HII galaxies (cf. Figure 10) this is resolved as the brightest
filament or a broken ragged shell sitting very close to the exciting cluster.
The central bright condensations
are  much smaller than the sizes of the  ionized volumes, which become 
more and more diluted at large radii to finally merge with the background galactic interstellar medium. 

The energy powering these giant volumes comes from the recently found unit of massive star formation: 
super star clusters (Ho 1997; see also Lamers et al. 2004 and references therein).
Massive concentrations of young stars within the range of 10$^5$ M$_\odot$ to several $10^6$ M$_\odot$, 
all within a small volume $\sim$ 3 -- 10 pc.
Multiple young super star clusters are found within the most luminous HII galaxies and 
starburst galaxies, as in M82 where a collection of two hundred of them 
have been found within the central 150 pc of the dwarf galaxy (Melo et al. 2005), while the most compact HII galaxies 
and particularly giant HII regions,
are structured by the action of only one or  a few of these exciting clusters.    

However, despite multiple efforts we still lack a detailed model. 
None of the numerical simulations in the literature, that have considered  either the powerful winds or the
impact of the ionizing radiation on the surrounding ISM, or both, have reproduced the kpc-scale  
morphology and inner structure of giant HII regions and 
HII galaxies. There are thus a number of central questions that remain in this field. And thus apart from the fact
that we ignore the initial condition, the distribution of matter around new massive bursts of star formation,  
it is crucial to understand how the central brightest condensations or ragged shells 
survive or withstand the full power of the exciting clusters, and how the multiple nested shells develop within the 
large-scale ionized volume and acquire their inner filamentary structure.  

Here we address all these issues by means of two dimensional (2-D) hydrodynamics. As in all other studies, our initial conditions are arbitrary 
and perhaps much more so in this case, as we have surrounded the central powerful SSC 
by a large concentration of spherical, static, cloudlets, reflecting perhaps 
the simple fact that we do not know the initial distribution of interstellar matter 
around a new major burst(s) of star formation. 
Our initial and boundary conditions are also arbitrary, in the sense that 
we approach the energetics from stellar clusters following the frequently used mode of instantaneous star formation
(see Leitherer \& Heckman 1995), although a continuous star formation spread over a few Myr may be a more reasonable approximation.
In section 2 we give a detailed description of the boundary and initial conditions used in our calculations.  Section 3 describes the results of 
four different cases in which we explore the impact that energetic winds or photoionization, or both, may have on  the selected cloudlet 
distribution and its surrounding gas. Our conclusions together with a full discussion of the results are given in section 4.

\section{Feedback from massive star clusters}

Superstellar clusters  are now recognized as the main unit of massive star formation in starburst galaxies (Ho 1997),
and their mechanical energy and radiative output the main feedback restructuring 
agents of the ISM (see Tenorio-Tagle et al. 2005).
Here we assume, in all cases,  a major central star formation burst
with a stellar mass  $M_{SC}$ (= $3 \times 10^6$ M$_\odot$) and a Salpeter IMF for stars between 1 and 100 M$_\odot$,
all concentrated within a radius $R_{SC}$ = 5 pc.

\subsection{The initial condition}

\begin{center}
\begin{figure}[htbp]
\vspace{17.0cm}
\caption{The initial condition. The  panel shows 
a cross-sectional cut along the computational grid showing 
isodensity contours with a separation $\Delta$ log $\rho$ = 0.18.
The central superstar cluster has a  power of $10^{41}$ erg s$^{-1}$ and 
a radius of 5 pc. Note that the velocity field within the central SSC 
grows from the sound speed value to its terminal velocity
$V_\infty$ =  10$^3$ km s$^{-1}$, here represented by the largest arrows. The 
plot displays one quadrant of the full cloudlet distribution within the central 50 pc $\times $ 50 pc of 
the total computational grid which spans 200 pc $\times$ 200 pc.}
\label{fig1}
\end{figure}
\end{center}

The mechanical energy and the UV photon output from the massive stellar cluster is here confronted with an ISM
structured into a collection of clouds. Three dimensional calculations are required if the clouds are spherical,
however, in this first 2-D approach to the problem each 3-D individual cloud is instead a torus
of which a 2-D cross-sectional cut is a disk, as depicted in Figure 1.   
Just as in other recent contributions (e.g. Pittard et al. 2005), 
to reduce the computational cost we restrict ourselves to 2-D simulations. 
Although we do expect some differences between 
2-D and 3-D calculations, clearly an important insight on the evolution and on the role played by the various parameters 
can still be obtained from the less computationally demanding 2-D simulations. 
Several two dimensional calculations using as initial condition  
the adiabatic SSC model of Chevalier \& Clegg (1985) have been performed with the 
explicit Eulerian finite difference code described 
by Tenorio-Tagle \& Mu\~noz-Tu\~n\'on (1997, 1998). This has been
adapted to allow for the continuous replenishment of matter within  the  SSC volume (see below).
The time dependent calculations  do not consider thermal conductivity but do account for radiative 
cooling (Raymond et al. 1976) for a gas with solar metallicity. 
Several calculations were made to reassure that the spatial resolution used led to  a convergent  solution.
All calculations here presented were made with the same numerical resolution of $\Delta R$ = $\Delta Z$ =  0.33 pc and
all of them with an open boundary condition along the grid outer edge.

Here we assume that most of the gas left over from star formation conforms a large collection of dense cloudlets  
around a central, massive ($M_{SC} = 3 \times 10^6$ M$_\odot$) starburst (see Figure 1). In the various cases here shown 
the only  difference is the cloudlet gas density ($n_c$ = 10$^2$ and 10$^3$ cm$^{-3}$).
The dense cloudlets are embedded and initially in pressure equilibrium with a low density 
medium ($n_{ic}$ = 0.1 cm$^{-3}$) that permeates the whole computational grid (200 $\times$ 200 pc).
In all cases the stationary clouds with a size, $r_{cl}$ = 7.7 $\times$ 10$^{18}$ cm,  have been placed at arbitrary locations (see Figure 1),
with an almost constant separation $\Delta R$ = 2 $\times$ 10$^{19}$ cm, $\Delta Z$ = 2.2 $\times$ 10$^{19}$ cm, except for the first two rows 
near the grid equator for  which the separation $\Delta Z$ = 2.4 $\times$ 10$^{19}$ cm. There are also a few smaller clouds scattered 
within the cloud stratum  placed there, as all other clouds, with the sole 
aim of obstructing the general radial outflow expected in an otherwise 
constant density case. 

\subsection{Boundary conditions}

In all cases we have  assumed that within the region 
that encompasses the recently formed stellar cluster ($R_{SC}$),
the matter ejected by strong stellar winds and supernova explosions is 
fully thermalized (Chevalier \& Clegg 1985; see also  Raga et al. 2001; Silich et al. 2004). 
This generates the large overpressure
responsible for the mechanical luminosity associated to the 
cluster which results from the  mechanical 
energy ($L_{SC}$) and mass (${\dot M}_{SC}$) deposition rates, 
within the star forming region. The total mass 
and energy deposition rates define the central temperature T$_{SC}$ ($\sim 1.5 \times 10^7$ K)
and thus the sound speed $c_{SC}$ ($\sim$ 500 km s$^{-1}$) at the cluster surface.

\begin{equation}
      \label{eq.01} 
T_{SC} = \frac{0.3 \mu}{k} \frac{L_{SC}}{{\dot M}_{SC}} , 
\end{equation} 

\noindent 
where $\mu$ is the mean mass per particle and $k$ the 
Boltzmann constant. On the other hand, the density of matter streaming from the SSC surface ($\rho_{SC}$)
can be obtained from the stationary condition in which the matter exiting the 
cluster surface ($R_{SC}$) per unit time, has to be equal to the mass deposition rate ($\dot M_{SC}$):

\begin{equation}
\dot M_{SC} = 4 \pi R_{SC}^2 \rho_{SC} c_{SC}.
\end{equation} 
The outflow starts its expansion 
 with its  sound speed ($c_{SC}$),  
however,
as it streams away it is immediately 
accelerated by the steep pressure gradient to rapidly reach
its terminal velocity ($V_{\infty} \sim 2 c_{SC}$ $\sim 10^3$ km s$^{-1}$). 
This is due to a fast 
conversion of thermal energy into kinetic energy of the resultant 
wind. 
Throughout the calculations, the density, temperature  and thus sound speed, are replenished within $R_{SC}$ at every time step.
In this way, as the wind develops, its density, temperature and thermal pressure 
approach their asymptotic values: $\sim r^{-2}$, $r^{-4/3}$ and  $r^{-10/3}$, respectively, 
while, the wind velocity reaches its maximum value ($V_{\infty})$.

\subsection{The time evolution} 

Cases A and B consider
the impact that each of the feedback agents, capable of structuring the ISM (SSC winds and photoionization),
may have independently on the selected cloudlet distribution. In both cases the
cloudlet gas  number density is $n_c$ = 10$^3$ cm$^{-3}$. 
The results shown in Figure 2 correspond to case A, that considers only a powerful SSC wind 
($L_{SC} = 10^{41}$ erg s$^{-1}$).
Figure 3 displays the other extreme situation
(case B) in which only photoionization 
has been  considered. Figure 2 shows density and temperature 
along a time sequence and the four panels in  Figure 3 
display only the run of density of the photoionized material (all at $T \sim 10^4 $ K) at selected evolutionary times. 
The color coded  figures indicate steep gradients
both in density and  in temperature, tracing strong compression and surfaces across which large  heating/cooling occurs
and thus indicate the location of shocks. 
Also noticeable in the plots are the sudden changes in velocity and 
direction of the various streams and thus the velocity field is also a good tracer of the 
presence of shock waves.

\subsection {Case A- The wind-cloud stratum  interaction}

\begin{figure}[htbp]
\vspace{19.0cm}
\caption{Case A. The initial interaction of the SSC wind with the cloud stratum. The panels display  
cross-sectional cuts along a section of the  computational grid showing: 
isodensity (left panels, in log units of g cm$^{-3}$) and temperature  (right-hand side panels, in log units of K).
The velocity field is also plotted in the density panels where
the longest arrow represents 10$^3$ km s$^{-1}$. The model is shown 
at three different times: 2.7 $\times 10^4$ yr, 8.8 $\times 10^4$ yr
and 1.83 $\times$ 10$^5$ yr, respectively. The size of the panels is 60 pc by 60 pc. 
}
\label{fig2}
\end{figure}

The powerful isotropic wind from the
central SSC immediately develops a leading shock moving with a velocity $V_S$. As this  
begins to  interact with the surrounding gas and as the free wind
approaches the cloud stratum a global reverse shock is established (see Figure 2). At the reverse shock the wind 
is thermalized, although initially  as the shape of the reverse shock follows closely the density or 
cloudlet stratification it acquires a multiple bow configuration. Consequently,
large azimuthal sections of the isotropic wind are only partly thermalized and
redirected by the bow sections of the reverse shock to stream around the encountered cloudlets (see the time sequence shown in Figure 2). 
The wind-cloudlet interaction leads to a complicated hydrodynamical set of events as it also drives a transmitted shock 
into the obstacle cloudlets.  The strength of this shock is proportional to the square root of the 
contrast density between the two media
($V_t = V_S (\rho_{ic}/\rho_c)^{1/2}$). Given the density contrast between clouds and the intercloud medium these disturbances
move slowly into the clouds and in this way,
the main  shock, rushing at first with its velocity $V_S$, rapidly 
circumvents the condensations while leaving a large pressure 
behind them. This pressure induces also secondary shocks into the overtaken cloudlets, what reduces   
their physical   cross-section. However, the strength of the secondary shocks is also
bound by the cloud-intercloud contrast density and, as in the case of the transmitted shocks, 
secondary shocks evolve also
slowly into the clouds.
The main shock and its overtaken matter, both driven by the partly thermalized wind, 
soon encounter other cloudlets and this induces, once again, the whole 
hydro response (reverse bow-shock into the wind and  transmitted and secondary disturbances into the cloudlets).

The partly thermalized wind behind the multiple bow shocks, plays an important role in the erosion of cloudlets (Tenorio-Tagle \& Rozyczka 1986;
Klein et al. 1994).
This is through  Kelvin-Helmholtz surface instabilities promoted by the rapid streams of shocked wind gas, that  continuously peel  the outer
skins of clouds (see density and temperature in the second and third panels of Figure 2). During the process, the matter ablated from the
clouds is driven into 
elongated streams, shaped by the multiple currents of
thermalized wind gas, while forming bridges into neighbouring clouds.

Clearly, as a result of all of these multiple interactions the wind is finally fully thermalized and  then 
streams to follow the leading shock into all possible paths of
least resistance in between the cold ($T \sim 10$ K) clouds. 
Eventually, say after 8.8 $\times 10^4 $ yr, the shocked wind and its leading shock finally begin to emerge 
at several places into the 
surrounding low density medium.  The further evolution, also shown in Figure 2 (third row), displays the multiple channels taken 
by the shocked wind through the cloud stratum as well as all of those that eventually
become blocked by matter ablated from clouds near the SSC. 
At the same time, clouds at the outer edge of the cloudlet distribution become ablated and conform large, almost radial streams 
of gas with a temperature $T \leq 10^4 $ K embedded into the hotter $T > 10^7$ K shocked wind gas (see last row of panels in Figure 2).

\subsection {Case B- Photoionization effects}

The effects induced by photoionization act in a completely different manner.
Here we assumed in all cases a large UV photon flux, able to 
establish a large-scale HII region with a size (a Str\"omgren radius) 
much  larger than the 
size of the computational grid. 
The hydrodynamical response to photoionization alone (shown in Figure 3) is also significant and is to compete 
with a central wind in the disruption of the dense stratum. 
Upon turning the cluster on, the temperature of the initially neutral 
medium is almost immediately set, through photoionization, to
the equilibrium temperature of the resultant HII region ($T_{HII}$ $\sim$ 10$^4$ K). 
This causes a large pressure imbalance  between the cloud and the intercloud medium which immediately 
promotes the expansion and dilution of the cloudlets  into the low density intercloud volume, through
a plethora of well localized "champagne flows" (see Tenorio-Tagle 1979).  As a result of such a champagne bath, the low density 
paths between cloudlets are to become narrower and narrower as time goes by.

\begin{figure}[htbp]
\vspace{19.0cm}
\caption{Case B. Photoionization effects.
The  panels show 
cross-sectional cuts along the computational grid showing 
isodensity values (in log units of g cm$^{-3}$) and 
the velocity field for which the longest arrow represents 45 km s$^{-1}$.
The size of the panels is 60 pc $\times$ 60 pc.
The evolutionary times are: The initial condition ($t$ = 0; upper left panel) and 
2.1 $\times$ 10$^5$ yr  (upper right-hand panel), 3.3 $\times$ 10$^5$ yr 
(lower left-hand panel) and 4.6 $\times$ 10$^5$ yr (bottom right-hand panel), respectively.
}
\label{fig3}
\end{figure}

The expansion of the dense stratum driven by the UV radiation, is aimed at 
restoring pressure equilibrium everywhere and as 
all the gas presents the same temperature, $T \sim T_{HII}$, this could only be reached if the fluid acquires  
an even density everywhere. This promotes a significant large-scale champagne outflow 
that rapidly composes an expanding halo, a growing rim of ionized gas around the cloudlet distribution, that ends up acquiring a
speed of the order of 40 km s$^{-1}$ (see Figure 3). Meanwhile within the cloud stratum, 
the expansion of clouds leads to multiple well localized converging flows that also rapidly build new condensations 
between the original cloudlets. The expansion of the latter also promotes converging flows that end up restoring, at least
partially, the original condensations, to then start once again their disruption. 

In the absence of a central wind, 
the central cavity is also filled with matter evaporated from condensations originally set close to the 
core of the cloud stratum (see the time sequence shown in Figure 3).

\subsection{Case C. Star-formation feedback into a high density cloudlet distribution}

There are thus two competing events:
The pressure acquired by the wind at the reverse shock ($P_{bubble} = \rho_w v_{\infty}^2$) 
defines the velocity that the leading shock
may have as it propagates into the intercloud medium ($V_S = (P_{bubble} / \rho_{ic})^{0.5}$), 
and thus, as a first approximation, if the cloudlet distribution has an extent 
$D_{cl}$ the time for the leading shock to travel across it is $t_{s}$ ($= D_{cl} / V_S$). On the other hand, 
the pressure gradient between ionized cloudlets and the intercloud medium, established through photoionization,
is to disrupt clouds and lead to a constant density medium in a time $t_{d} = \alpha d_{cs} / (2 c_{HII})$;
where $d_{cs}/2$ is half the average distance or separation between 
clouds, $c_{HII}$  is the sound velocity in the ionized medium 
and $\alpha$ is a small number ($\sim$ 4 - 6) and accounts for the number of times that a 
rarefaction wave ought to travel (at the sound speed) the 
distance $d_{cs}/2$ to replenish the whole volume with an average even
density $<\rho>$. In this way if the cloud disruption process 
promoted by photoionization leads to an average $<\rho>$ that 
could largely reduce the velocity of the leading shock 
($V_S = (P_{bubble} / <\rho>)^{0.5}$), then $t_{s}$ could be larger 
than  $t_{d}$, leading, as shown below, to an effective
confinement of the shocked wind, at least for a significant part of the evolution.

\begin{figure}[htbp]
\vspace{17.0cm}
\caption{
The star formation feedback into a dense cloudlet distribution. 
Case C, $n_c$ = $10^3$ cm$^{-3}$, $L_{SC} = 10^{41}$ erg s$^{-1}$.
The panels display  
cross-sectional cuts along a section of the  computational grid showing: 
density (left panels) and  temperature (right-hand side panels)
together with the velocity field 
for which the longest arrow represents 10$^3$ km s$^{-1}$. The evolution time (from top to bottom panels)
is: 7 $\times 10^4$ yr, 1.25 $\times 10^5$ yr
and 3.26 $\times$ 10$^5$ yr, respectively.  
The figure shows the  60 pc $\times$ 60 pc central section  of the computational grid.
}
\label{fig4}
\end{figure}

Case C considers both feedback events: a powerful SSC wind with a power of 10$^{41}$ erg s$^{-1}$ and a 
ionizing photon flux (10$^{54}$ UV photons s$^{-1}$) sufficient to photoionize initially all the matter within the computational grid. 
Figure 4 displays the rapid evolution of the photoionized cloud stratum,
filling almost everywhere the low density intercloud zones, while 
spreading the density of the outermost cloudlets into the surrounding intercloud medium.
This causes the development of an increasingly larger ionized expanding  rim around the cloudlet distribution
and  leads to a rapid enhancement of the intercloud density. 
The supersonic expansion of the champagne gas 
can better be traced by the multiple isothermal shocks, apparent  in the temperature plots 
(these are depicted as narrow blue lines in the  first and second temperature panels in Figure 4).
Noticeable also in Figure 4, is the location and shape of the 
global reverse shock into the wind, that evolves from  an initially ragged surface across which the 
isotropic wind is only partly thermalized, to an almost hemispherical surface that fully thermalizes the wind. 
Initially, after crossing the reverse shock
the hot wind drives the leading shock into the cloudlet distribution and this immediately looks for
all possible paths of least resistance in between clouds. This fact, as shown in the density and temperature panels in Figure 4,
diverts the shocked wind into multiple streams behind every overtaken cloudlet,   
diminishing steadily its power  
to reach the end of the cloudlet distribution before the outermost clouds expand and block many of the possible exits
into the low density background gas. 
Note that in the absence of clouds, the wind traveling at 10$^3$ km s$^{-1}$, 
would have  taken 2$ \times 10^5$ yr to get to the edge of the computational grid. 
In the calculation only one channel, the initially widest channel close to the grid equatorial axis, 
across which the original cloudlet spacing was set 
slightly larger than in the rest of the distribution, is succesfully crossed by the 
leading shock and the thermalized wind behind it, after $\sim$ 10$^5$ yr of evolution (see second row of panels in Figure 4). 
A second channel through the cloud stratum is completed 
(close to the symmetry axis) after a time 
$t = 5 \times 10^5$ yr (see first and second rows of panels in Figure 5). 
All other possibilities, evident for example in the second frames of Figure 2, are here blocked by the 
large densities resultant from the champagne bath (compare Figure 4 with Figure 2).

\begin{figure}[htbp]
\vspace{17.0cm}
\caption{The time evolution of case C, as in Figure 4, density and temperature plotted 
at three different times (from top to bottom panels): 4.6 $\times 10^5$ yr, 6.32 $\times 10^5$ yr
and 8.16  $\times$ 10$^5$ yr, respectively.  The panels show the full computational grid (200 pc $\times$ 200 pc).
}
\label{fig5}
\end{figure}

Figure 5 displays details of the resultant 
structure after 4.6 $\times 10^5$ yr of evolution. At this time, 
a bunch of partly shocked cloudlets and most of the extended outer rim of photoionized gas
obstruct  still the direct expansion of the thermalized wind into the background low density ISM. 
Matter able to emit at optical wavelengths is depicted in the figures at
temperatures  in the range $T \sim 10^4 - 10^5$ K. Within this temperature range is the low density background gas and 
the ablated cloudlet gas driven into an almost hemispherical elongated ragged shell trapped between the shocked wind and the 
extended  photoionized rim. 
More gas at this temperature range is found at the outer boundary of the 
large-scale expanding shells, as well  as in the initial sections of the elongated filaments that extend from the original
cloudlet distribution well into the interior of the  large-scale bubbles. The filaments result from cloud ablation, and delineate 
at least initially, the open channels across which the
streams of thermalized wind matter exit the cloudlet distribution. The rapid streams across the open channels thus 
promote the build up  of large-scale shells
with a significant inner filamentary structure.

Figure 5 displays also the total extent, shape and evolution of the resultant filaments.
Indeed, the remains of ablated cloudlets can span 
up to hundreds of pc away from their original location. The continuous drag imposed by the rapid shocked wind streams, 
shock and heat the ablated gas up to  temperatures similar to those of the hot thermalized wind matter,
while the ablated cloud parcels acquire the speed of the streams.
As time proceeds, radiative cooling allows the thin large-scale outer shells
and the filamentary inner structure to become more prominent within the temperature range $10^4 \leq T \leq 10^5$ K
causing it to become more apparent  at optical wavelengths, while most of the volume 
remains hot ($T \geq 10^6$ K) and detectable  only in the X-ray regime.    

Clear in all panels, with the help of the
velocity field, is also the location of the 
global reverse shock that thermalizes the central wind. 
Most of the wind energy, throughout the calculation, is  radiated away (see below) 
while the shocked wind confronts the collection of cloudlets driven to compose the  dense ragged slowly moving
shell that covers and blocks most of the SSC sky.  
The strongly decelerated hot wind gas however,
meets its sonic point close to the open channels to stream supersonically and later be  partly redirected 
by multiple oblique and crossing shocks.  
Only a small fraction of the SSC mechanical luminosity is channeled into the giant shells and this explains their 
slowness to fill the computational grid, at  the same time that
multiple filaments enhance the structure within the giant bubbles.

The last calculated model (at $t = 8.16 \times 10^5$ yr; third row in  Figure 5) shows the moment of merging of 
the giant shells while these
have grown to exceed the dimensions of the computational grid. 
At the center of the plot one can see the well developed free-wind region bound
by the reverse
inner shock. The shocked  wind occupies now a broader zone ($\sim$ 15 pc in radial extent) and  is bound by
the cloud matter now composing  the densest (brightest) and slowly 
expanding ragged shell that has just began to move  
into the photoionized outer rim. All of this structure is now surrounded by the large-scale 
merging superbubbles, at the edge of which the swept up matter collapses into multiple thin outer shells.

\subsection{Case D. Star-formation feedback into a low density cloudlet distribution }

\begin{figure}[htbp]
\vspace{17.0cm}
\caption{The time evolution. Case D, $n_c$ = $10^2$ cm$^{-3}$, $L_{SC} = 10^{41}$ erg s$^{-1}$. 
The same as Figure 4 
at three different times: 8.7 $\times 10^4$ yr, 1.82 $\times 10^5$ yr
and 2.9  $\times$ 10$^5$ yr, respectively. 
The figures shows the inner  60 $\times $ 60 pc section of the  computational grid.}
\label{fig6}
\end{figure}

Here we show the evolution that results from a low density cloudlet distribution ($n_c = 100$ cm $^{-3}$) 
being impacted by the energetics from
a massive $3 \times $10$^6$ M$_\odot$ cluster. The input is  $L_{SC} = 10^{41}$ erg s$^{-1}$ and 3 $\times$ 
10$^{53}$ ionizing photons per second.
The latter causes  throughout the calculation  an HII region with a size larger than the computational grid. 
The results (in Figures 6 and 7), are to be compared with those of case C, shown in Figures 4 and 5, shown at 
the same physical scales. The evolution of this case reproduces all features found in case C although it
proceeds much faster. Clearly, 
the mass evaporated from the clouds through the multiple champagne flows, enhances the density and the thermal pressure everywhere 
and causes, as in case C,  
that the cloud stratus as a whole survives much longer further downstream. However in this case, the initial contrast 
density between clouds and the intercloud gas is so small that it allows the  
wind to become  dominant from the very start of the calculation. 
Figure 6 shows how the shocked wind 
rushes through the cloudlet distribution and is able to find many of the  possible channels much earlier 
than in the case of a denser cloud stratum. 
It is also able to rapidly store the clouds into an elongated and ragged slowly moving shell. This, as in the former case,  
inhibits, at least for a significant fraction of the evolution, the direct propagation of the shocked wind  
into the low density background gas. However        
in this case, the ragged shell that contains most of the ablated clouds collected together by the action of the shocked wind, 
becomes unstable towards the end of the calculation (see Figure 7) 
as it begins to  move into the extended photoionized rim that developed earlier around the original cloud stratum.  
The ragged shell is in this way disrupted into a collection 
of fragments  that immediately become exposed to the rapid streams that develop as 
the inner reverse shock acquires once again a multiple bow shock configuration. Meanwhile, the ablation of the resultant fragments provides
the large-scale superbubbles with an enhanced inner filamentary structure (see Figure 7).

\begin{figure}[htbp]
\vspace{17.0cm}
\caption{The time evolution. Case D, $n_c$ = $10^2$ cm$^{-3}$, $L_{SC} = 10^{41}$ erg s$^{-1}$. 
The same as Figure 5 at three different times: 2.9 $\times 10^5$ yr, 3.3 $\times 10^5$ yr
and 4.4  $\times$ 10$^5$ yr, respectively.  The panels show the full computational grid.
}
\label{fig7}
\end{figure}

\subsection{Summary}

\begin{figure}[htbp]
\begin{center}
\includegraphics[width=0.4\textwidth]{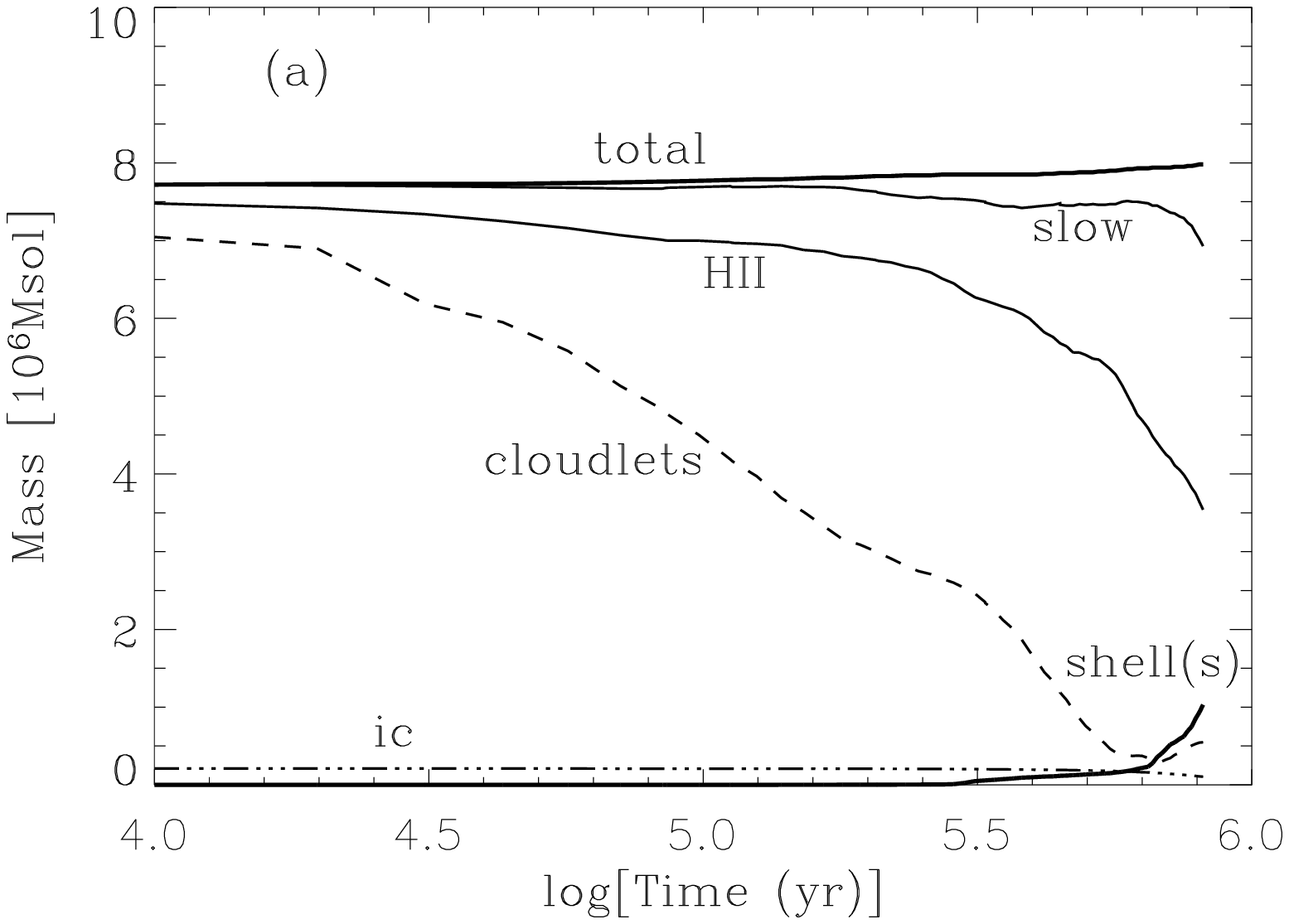}
\includegraphics[width=0.4\textwidth]{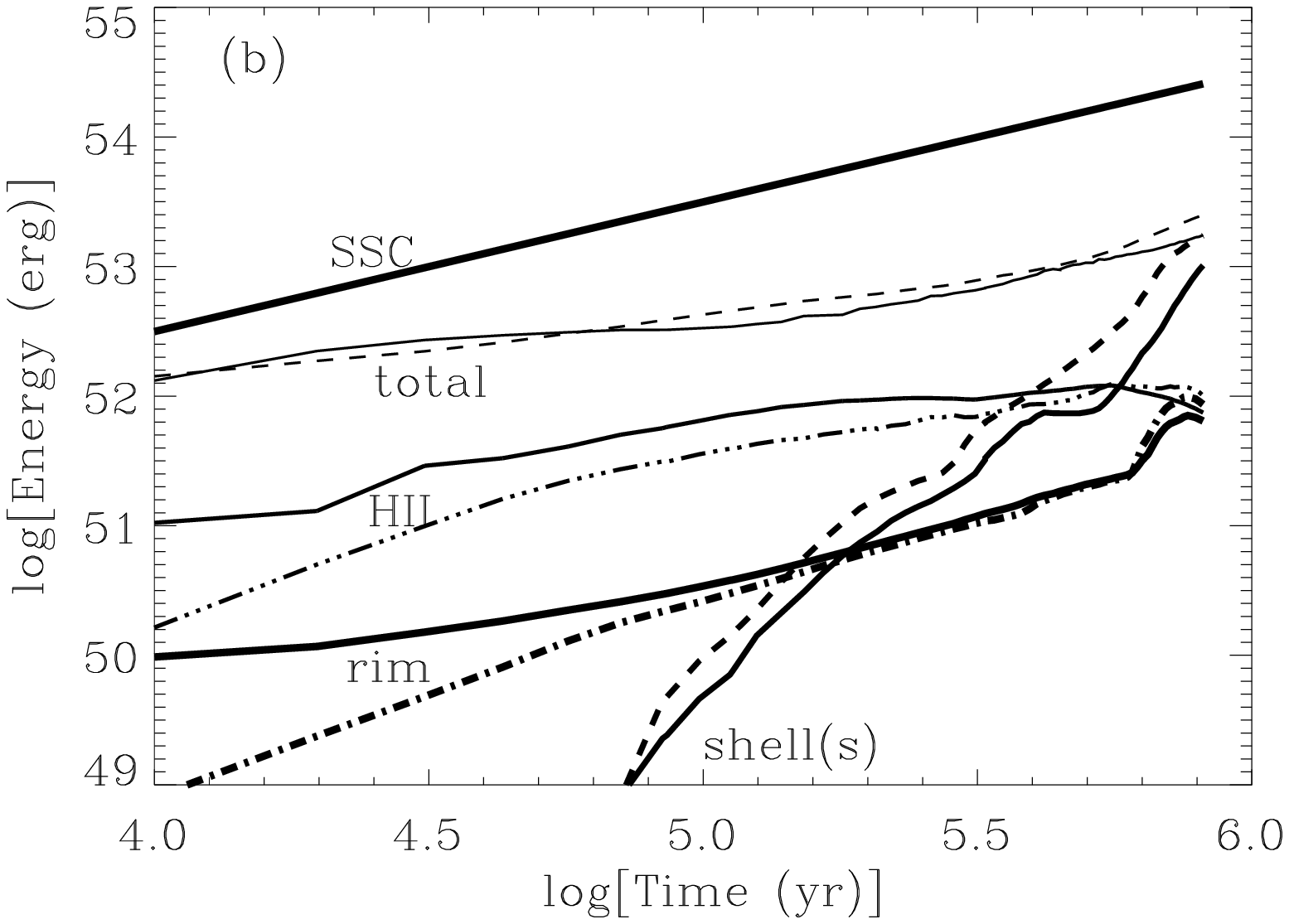} \\
\end{center}
\begin{center}
\includegraphics[width=0.4\textwidth]{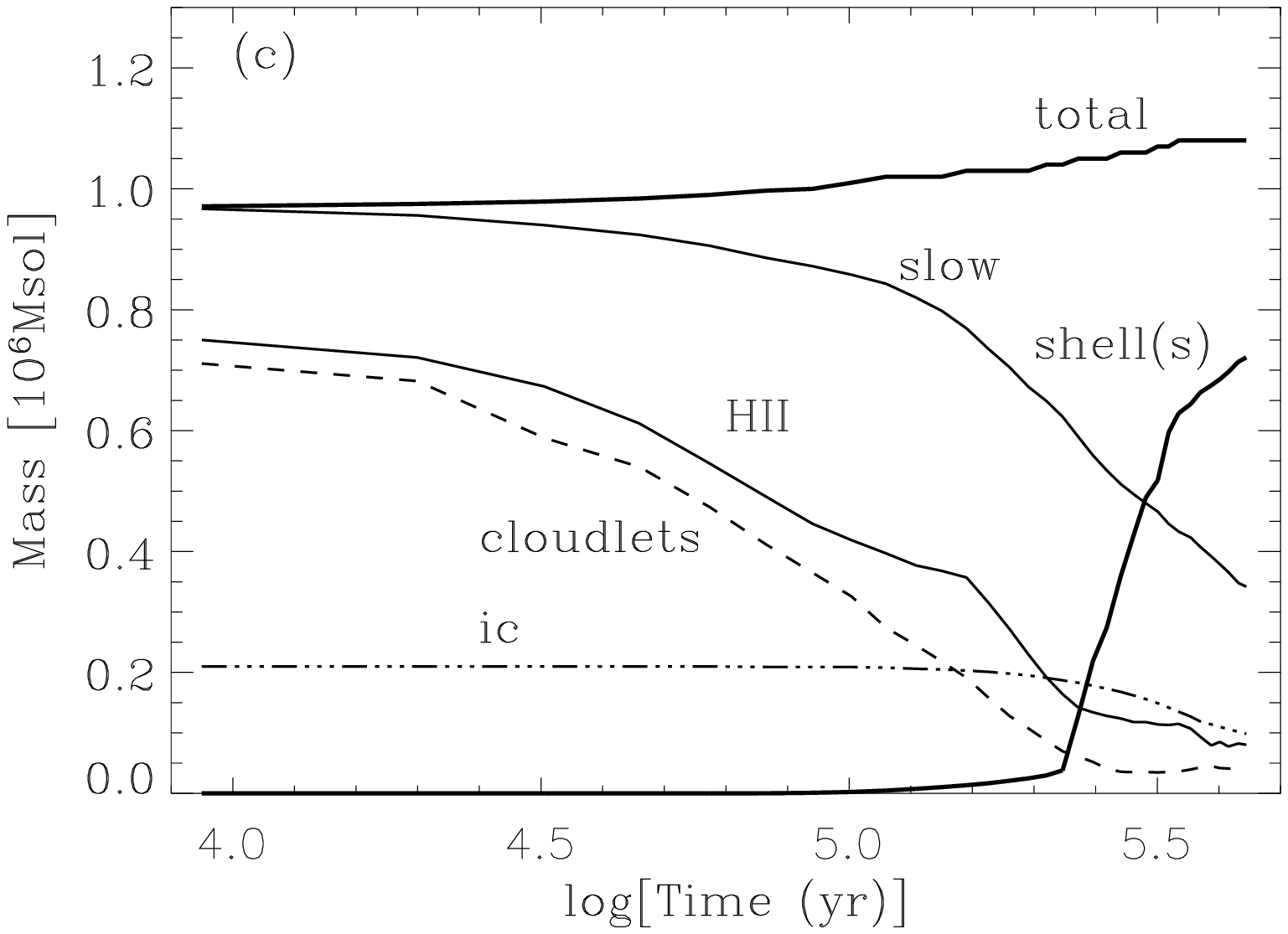}
\includegraphics[width=0.4\textwidth]{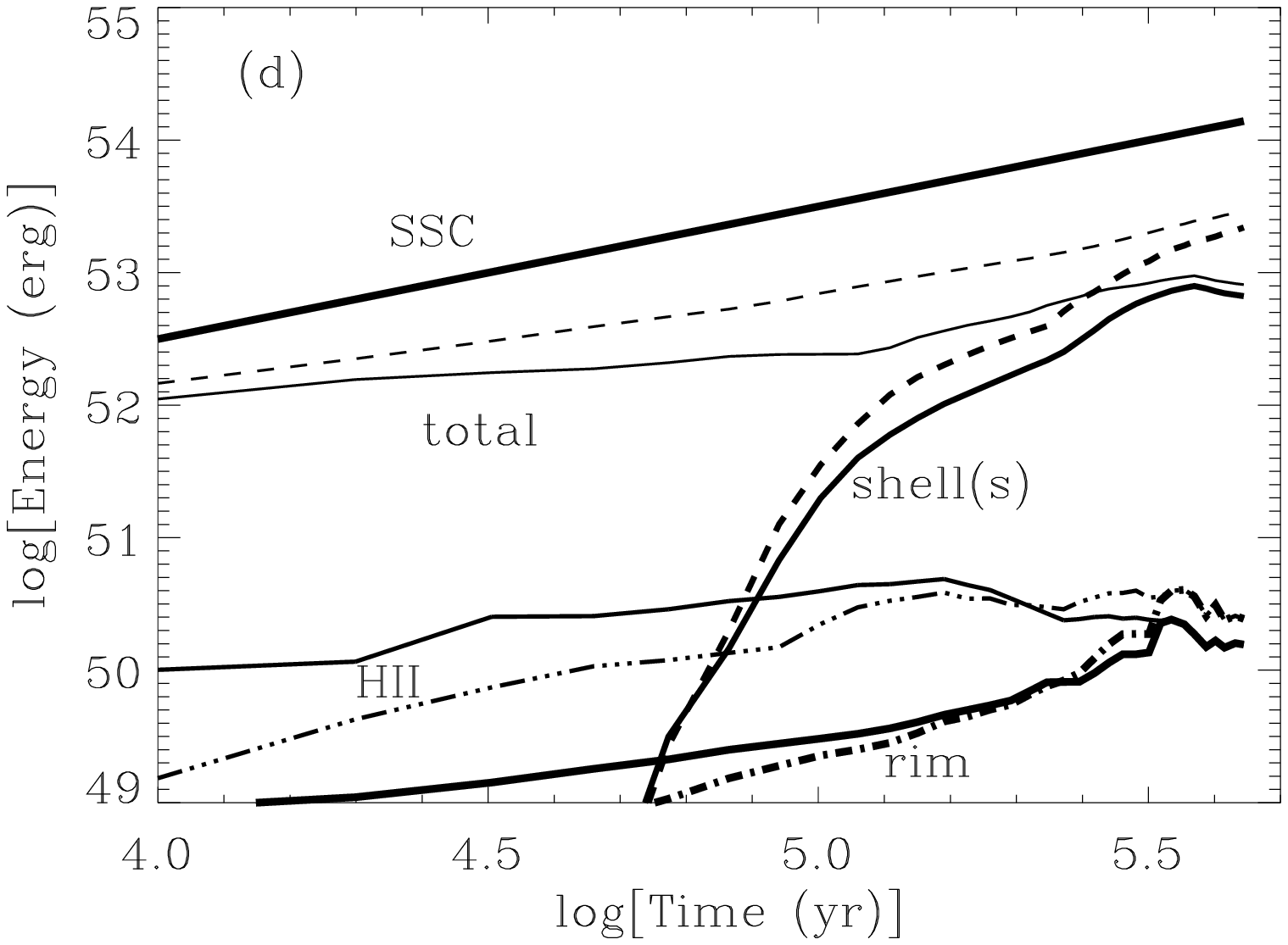} \\
\end{center}
\caption[EnergyBudget]{The time evolution. Mass and energy budget for the high density case C (upper panels) and the low density case D (lower panel).
The various lines in figures a and c represent: Total mass within the computational grid, matter moving with less than 50 km s$^{-1}$,
cloud (cloudlets) and intercloud (ic) matter that preserve their original density value, and HII and shells represent all the matter set into motion 
through a champagne flow and that has left the original cloudlet distribution volume, respectively. Panels b and d display the total amount of 
mechanical energy provided by the central cluster (SSC), the total kinetic (dashed line) and thermal (solid line) energy
found within the computational grid. The other curves marked HII, rim and shells, display the kinetic (broken lines) and thermal (solid lines) energy
stored in the ionized medium, or in the rim, or within the giant superbubbles.
}
\label{fig:shell}
\end{figure}

As a summary of the calculation, Figure 8 shows the mass and energy  budgets, 
as a function of time for cases C and D. The various lines in Figure 8a and c represent, from top to bottom: the total mass 
in the computational grid, which grows through the SSC mass deposition rate (labelled ``total'' in the figures).
Most of the mass in the grid  moves with a small velocity ($<$ 50 km s$^{-1}$).  This  includes all the photoionized gas as well as
the slowly moving cloud matter overtaken by the leading shock (marked "slow" in the figures).  The cloudlets and intercloud
medium (ic) lines represent the gas with zero velocity within the grid, that preserve their original density values, 
both of these are largerly 
reduced as a function of time.
The ``HII'' curve accounts for the photoionized gas set into motion through champagne flows. The final curve (``shells''), 
that appears towards the end of the calculations, accounts for all matter that has managed to stream out of 
the original cloudlet distribution. Figures 8b and d consider the energetics as a function of time of cases C and D, respectively. 
From top to bottom
the various curves show: The total amount of kinetic energy delivered by the central star cluster (labelled ``SSC''). 
The total amount of thermal (solid line) and kinetic energy (dashed line) within the 
computational grid (marked "total" in the figure). A comparison of the energy delivered by the central cluster with 
the addition of the latter two energies (total), shows that almost 90$\%$ of the SSC mechanical energy is radiated away 
throughout the evolution of case C, and 70$\%$ in case D. The remaining energy ends up being  stored in the gas that has managed to 
stream out of the original cloudlet distribution (labelled ``shells''). 
 The other curves marked ``HII'', ``rim'' and ``shells'', indicate the thermal (solid lines) and
kinetic energy (broken lines) stored in the ionized medium, part of which is in the photoionized outer rim, and finally the 
energy content within the large-scale expanding superbubbles, including the large-scale shells. 
Clearly, the latter amounts only to a small fraction of the energy delivered by the central SSC.

Figure 9 compares the size acquired  by the various disturbances that develop in case C, within the flow. 
We have measured the location of the photoionized rim outer boundary, as well as the 
position of the ragged shell and of the reverse shock as a function of time. Also the dimensions of the large-scale superbubble that develops
earlier in the calculation,  measured 
from the point where energy is fed into it, at the edge of the channel carved through the cloudlet distribution (see Figure 9). 
The trend followed by the expanding photoionized rim is initially in  excellent agreement with a champagne flow expansion speed 
of about 50 km s$^{-1}$ (see Franco et al 1990), and is strongly decelerated (to 30 km s$^{-1}$) in the late part of the evolution as the 
giant superbubble wraps around it (see Figure 5). The       
trends followed by the other disturbances have been  compared with the analytical formulation of a 
strong wind powered by a constant energy input rate and 
evolving into a constant density medium (see eg. Koo \& McKee 1992). From the formulation of Koo \& McKee we know that the reverse  shock
position  is determined by: $R_{rs} = 0.77 (L_{SC}/\rho_0)^{0.3} V_\infty^{-1/2} t^{0.4}$; where $\rho_0$ is the background density.
As shown in Figure 9, an excellent agreement is reached if one assumes the full power of the wind
($10^{41}$ erg s$^{-1}$) and  a value of $\rho_0$ = 100 cm$^{-3}\times m_H$ instead of the cloudlet density ($n_c$ = 1000 cm$^{-3}$), 
which we justify by the uneven density within the cloud stratum and
particularly 
by the low density values assumed for the  intercloud medium. Assuming the same value of $\rho_0$ for the leading shock, the expression 
$R_{ls} = 0.86  (L_{SC}/\rho_0)^{0.2} t^{0.6}$, of Koo \& McKee, leads to a disagreement of a factor of two with the location of the ragged shell. 
This is due to the increasingly larger loss of energy through the channels built through the cloud stratum,
which enhance their width as the evolution proceeds. This increasingly larger loss of energy is in fact fed into the large-scale superbubble and
thus its leading shock instead of decelerating and causing a trend $R_{ls} \propto t^{0.6}$ (expected for a constant energy input rate)
gives a trend $R_{ls} \propto t^{1.2}$, see Figure 9. 
This also implies that the energy input rate into the superbubble ($L_{SC} \propto t^{\alpha}$,  grows as $t^3$, consistent with the shell
energy budget in Figure 8b ($\alpha = log((E(t_2) - E(t_1))/E(t_1) + 1) / log(t_2/t_1) \approx 3.2$ if one considers the shell evolution 
between 10$^5$ yr and 3 $\times 10^5$ yr).
 
Further properties of the brightest filament or ragged shell may be derived from a consideration of the ram pressure exerted by the central wind ($\rho_w V_{\infty}^2$).
From the wind mechanical energy ($\dot E = 1/2 \dot M V_{\infty}^2$) and the continuity equation that establishes that $\dot M = 4 \pi R^2 \rho_w V_\infty$,
one can derive an expression for $\rho_w = \dot E/(2 \pi R^2 V_\infty^3$) and thus the wind ram pressure  $\rho_w V_\infty^2 =  \dot E/(2 \pi R^2 V_\infty$).
This is to be compared with the ragged shell ram pressure ($\rho_{rs} v_{rs}^2$) at different evolutionary times or considering  different values of $R$.
The expression is in good agreement with our calculations, leading to average
number density values of the ragged shell of a few thousand particles per 
cm$^3$ at the start of the evolution (for an $R \sim $ 10 pc) which fall to 
several hundred particles per cm$^{3}$ when $R \sim$ 50 pc.

The above estimate can be directly applied to well known sources. Such is the case of the brightest shell in NGC 604, for which the average distance to the 
exciting cluster is 30 pc (Gonz\'alez Delgado \& P\'erez 2000) and the inferred expansion speed amounts to 35 km s$^{-1}$ (cf. figures 3 and 4 in Tenorio-Tagle et al. 2000), 
leading to an average density of $\lessapprox100$ cm$^{-3}$ (the exact value depending on the luminosity of the 
best fitted model, cf. table 4 in Gonz\'alez Delgado \& P\'erez 2000) in agreement with the
data (cf. figure 10 of Ma\'\i z Apell\'aniz et al. 2004). The agreement however leads also to a controversy. The dynamical time for such a shell ($t_{dyn}$ $\sim$ 10$^6$ yr) is smaller than the life time
estimated ($\sim 3 \times 10^6$ yr) for the exciting stars (Gonz\'alez Delgado \& P\'erez 2000). Possible solutions of such a puzzle are: a) Clusters born in low metallicity galaxies
are to undergo much milder winds before the supernova era starts ($\sim$ 3 Myr; see Leitherer \& Heckman 1995), allowing for a better match between the dynamical
time and the stellar lifetime. b) Also, a dense cocoon around the newly formed
sources may delay the impact of winds and photoionization in the background 
media, bringing the two times considered here into better agreement. 
Once the full energy from sequential supernovae and the displacement of
the dense cocoon surrounding the newly formed cluster are overcome, the giant structure of HII regions and HII galaxies would develop within a few 10$^6$ yr while
the giant shells acquire kpc dimensions and the ragged shell is displaced to a few tens of pc away from the central source. At the same time that the production of
UV radiation would steadely drop making less easy the detection of the nebular structure at optical wavelengths.

\begin{figure}[htbp]
\vspace{17.0cm}
\includegraphics{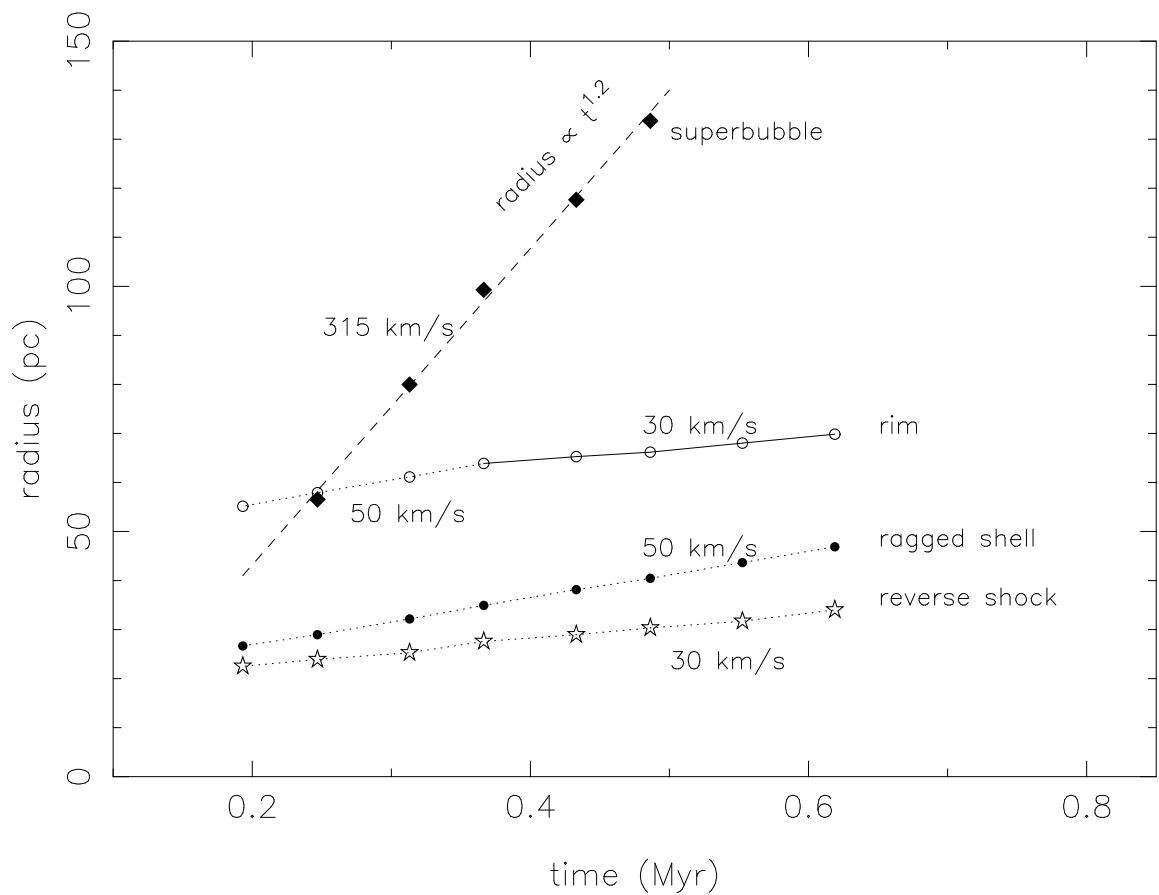}
\caption{The time evolution. The location of the various fronts that develop within the flow for case C, as a function of time. 
The figure indicates the location of the champagne rim outer edge, the ragged shell, the reverse shock and the furthest edge of
the giant superbubble. All of them with an indication of their expansion speed. The radius of the giant bubble has been fitted with 
a power law in which $r \propto t^{1.2}$.
}
\label{fig9}
\end{figure}

\section{Discussion}


We have shown here the effects of feedback from a  massive stellar cluster into a selected cloudlet 
distribution, which although arbitrary, as it could have had a different extent 
or it could have considered clouds of different sizes,
separations and locations, it has allowed us to explore a wide range of possibilities. These however, have lead  
to the two main 
possible physical solutions to be expected: partial pressure confinement of the shocked wind, 
and a stable and long-lasting filtering of the thermalized wind 
through the cloudlet stratum. Similar solutions would have been found 
for different  cloud densities and cloud distributions and/or for different SSC masses ($M_{SC}$) or energetics.
Thus despite the arbitrary and simple boundary and initial conditions  
here assumed, the structures that develop within the flow resemble  the structure of
giant HII regions and HII galaxies (see Figure 10).
In particular we refer to the giant and multiple well structured shells evident at optical wavelengths, often referred to as nested shells 
(see Chu \& Kenniccutt 1994), 
that enclose a hot X-ray emitting gas, as in 30 Dor (Wang 1999) and NGC 604 (Ma\'\i{}z-Apell\'aniz et al. 2004). 
The calculations also lead to the slowly expanding brightest filament (or ragged shell) 
present in all sources close and around the exciting stars, and to the elongated, 
although much fainter,  filaments on either side of the 
open channels that reseamble the ends of elongated columns into the giant superbubbles.
 
None of these structural features appear in calculations 
that assume a constant density ISM, nor in those that allow for a constant density molecular cloud as birth
place of the exciting sources, or in calculations where the sources are embedded in a plane stratified background atmosphere.
For all of these features to appear, a clumpy circumstellar medium seems to be a necessary requirement. A medium that would
refrain the wind from an immediate exit into the surrounding gas and that 
would also allow  for the build up of multiple channels through 
which the wind energy would flow in a less unimpedded manner. And thus 
giant HII regions and HII galaxies, both powered by massive star formation events producing  an ample supply of UV 
photons and a powerful wind mechanical energy, all of them seem to process their energy into a stratum of dense cloudlets 
sitting in the immediate vicinity of the star formation event.

\begin{figure}[htbp]
\vspace{17.0cm}
\caption{The observational evidence. HST H$\alpha $ images of the IIZw 40, the central regions of NGC 4214, and NGC 604 
(in M33), and an ESO NTT image of 30 Doradus, all of them ploted in a linear parsec scale. For IIZw 40 the inset shows the most central region.
HST images were retrieved from the HST archive at STScI. The NTT image courtesy of Jes\'us 
Ma\'\i z-Apell\'aniz and Rodolfo Barb\'a.
}
\label{fig-ha}
\end{figure}

Our calculations confirm many of the results by Pittard et al. (2005)
for the rapid sequence of the hydrodynamical events that emerge from the wind-cloud early interaction. In particular, a global 
bow shock is to engulf two or more cloudlets if the separation between them is smaller than their diameter 
while  individual bow shocks are set around each condensation when they are further appart from each other, as shown in 
all small-scale Figures 2, 4 and 6.
Also, the elongated tails of ablated matter behind bow shocks, are not necessarily aligned with the incoming stream velocity vector. 
The deviation in the alignment of tails is due to pressure gradients on either side of each cloud. As shown also 
by Pittard et al. (2005) and in 
our small-scale figures, tails that show deviations from the alignment with the incoming wind end up pointing 
away from the cloudlets, and when the flow is subsonic the tails end up pointing towards each other.  Here we have shown 
that the ablated matter that composes the various tails, ends up merging with other cloudlets further downstream to eventually
compose the almost even density, slow moving ragged shell. 

Given the large number of possibilities regarding SSCs, we consider  that 
cases A and B also seem relevant. One can imagine situations  in which the stellar photon output is
more developed and dominant than the winds, as in the initial stages of the evolution of coeval clusters, and thus
leading to results similar to those of case B. Also cases in which the photon flux 
has dropped sufficiently, as predicted for coeval clusters after 3 Myr of evolution, leading (as in case A) to a long-lasting SSC wind
as the dominant feedback mechanism.

In the calculations here shown,
as the selected cloudlet stratum covers most of the sky above the central SSC, all calculated cases follow a similar evolutionary track
in which two competing hydrodynamical processes define the time evolution. On the one hand, photoionization that aims at establishing a 
constant density medium everywhere and, on the other, the wind mechanical energy that looks for all 
possible exits out of the cloudlet distribution.
In all cases that consider both effects, the ensemble of clouds ends up being driven to compose an almost constant density, 
low velocity ragged shell, that acts as
a barrier to most of the shocked wind.  In this way the wind matter becomes subsonic upon crossing 
the reverse shock and trapped behind the ragged shell. On the other hand, 
channels through the cloudlet stratum allow for the rapid streaming of the shocked wind gas and lead to the build up of large-scale
superbubbles with a pronounced  inner filamentary structure. 
The latter is promoted by the ablation of clouds from each of the channels inner edges.
The calculations also show that for individual giant bubbles to be easily recognized they would have to be fed through channels 
in the cloud stratum, carved well further apart from each other. 
Neighboring channels would rapidly end up bursting their carried  shocked wind energy into a single shell.  

Photoionization tends to equalize the density everywhere, causing an evolution towards a higher even pressure, 
however its power to slow the effects of the wind, its power to make the wind less dominant strongly depend
on the initial contrast density between the cloud and intercloud medium. And thus, the larger the initial contrast density, 
the less dominant and longer lasting the impact of the wind would be.

\acknowledgments

We are grateful to an anonymous referee for useful suggestions that 
have led to an improvement of this work.

GTT acknowledges finantial support from the Secretar\'\i{}a de Estado 
de Universidades e Investigaci\'on (Espa\~na) through grant
SAB2004-0189 and the hospitality of the Instituto de Astrof\'\i{}sica 
de Andaluc\'\i{}a (IAA, CSIC) in Granada, Spain. This study has been 
partly supported by grants AYA2004-08260-C03-01 and AYA 2004-02703
from the Spanish Ministerio de Educaci\'on y Ciencia, grant TIC-114 
from Junta de Andaluc\'\i a and Conacyt (M\'exico) grant 47534-F.

This research has made use of NASA's Astrophysics Data System 
Bibliographic Services. Some of the data presented in this paper were 
obtained from the Multimission Archive at the Space Telescope 
Science Institute (MAST). STScI is operated by the Association 
of Universities for Research in Astronomy, Inc., under NASA contract 
NAS5-26555. Support for MAST for non-HST data is provided by the 
NASA Office of Space Science via grant NAG5-7584 and by other grants 
and contracts. We have made use of STSDAS and PyRAF which are 
products of the Space Telescope Science Institute, which is operated 
by AURA for NASA. IRAF is distributed by the National Optical 
Astronomy Observatories, which are operated by the Association of 
Universities for Research in Astronomy, Inc., under cooperative 
agreement with the National Science Foundation.

\end{document}